\documentstyle[stwol]{article}

\input{psfig}

\def\be{\begin{equation}}
\def\ee{\end{equation}}
\def\bea{\begin{eqnarray}}
\def\eea{\end{eqnarray}}

\begin{document}

\title{QCD DIPOLE PREDICTION FOR DIS AND DIFFRACTIVE STRUCTURE 
FUNCTIONS }

\author{ Ch.ROYON }

\address{DAPNIA-SPP, Centre d'Etudes de Saclay, F-91 191 Gif-sur-Yvette Cedex,
France}


\twocolumn[\maketitle\abstracts{The $F_{2}$, $F_{G}$, $R=F_{L}/F_{T}$ 
proton structure
functions are derived in the QCD dipole picture of BFKL dynamics. 
We get a three parameter fit describing the 1994 H1
proton structure function $F_{2}$ data in the low $x$, moderate $Q^{2}$
range. Without any additional parameter, the gluon density and the
longitudinal structure functions are predicted.The diffractive dissociation
processes are also discussed, and a new prediction for the proton diffractive
structure function is obtained. 
\footnote{Invited
talk given at the 28th International Conference
on High Energy Physics (ICHEP 96), 25-31 July 1996, Warsaw, Poland}
}]



The purpose of this contribution is to show that the QCD dipole picture
\cite{Muel+}
which contains the BFKL dynamics provides a pertinent model for describing
the proton structure function at HERA in the low $x$ and moderate $Q^{2}$
range, as well as the diffractive structure
function. The recently published H1 data might provide an opportunity to
distinguish between the different QCD evolution equations
(DGLAP and BFKL equations
\cite{equ})
for small $x$ physics. This is why it is very important to test the
BFKL dynamics using the dipole model and $k_{T}$ factorisation
\cite{catani}. We
can then get predictions for $F_{2}$, $F_{G}$, and $R=F_{L}/F_{T}$ proton
structure functions, and for the proton diffractive structure functions
which represents a new prediction of the dipole model.

\section{BFKL dynamics in the QCD dipole framework}
To obtain the structure function $F_{2}$, we use the $k_{T}$ factorisation
theorem which is valid at high energy (small $x$). In a first step,
we calculate the deep inelastic cross section $\sigma^{onium}$ of
a photon of virtuality $Q^{2}$ on an onium state
(heavy $q \bar{q}$ state).
This onium state can be described by  dipole configurations \cite{Muel+}.
The photon-onium cross-section reads
$\sigma^{onium} = \int^{ }_{ } d^2 r d z \Phi^{(0)} (r,z) \sigma(x,Q^2;r)$
where $\Phi^{(0)}$ is the probability distribution of dipole
configurations of transverse coordinate $r$.
In the $k_T$-factorization scheme \cite{catani}, one writes
\begin{eqnarray}
\label{fact}
Q^2 \sigma (x,Q^2;r) = & \\  \int^{Q^2}_{ } d^2 \vec{k} \int^{1}_{0}
\frac{d z}{z} \nonumber  
& \hat{\sigma} (x/z,\vec{k}^2 / Q^2) \, F(z,(kr)^2)
\end{eqnarray}
where $\hat{\sigma}/Q^2$ is the $(\gamma~ g(k) \rightarrow q ~ \bar{q})$ Born 
cross section for an off-shell gluon of transverse momentum
$\vec{k}$. $F(z,(kr)^2)$ is the unintegrated gluon distribution of an onium 
state of size $r$  and contains the physics of the BFKL pomeron.
\par
Using once more the $k_{T}$ factorisation in extracting a gluon from a dipole
of transverse radius $r$ and assuming renormalisation group properties of $F$, 
one can show that \cite{ourpap}:
\begin{eqnarray}
\label{crosshadronnonpert}
& F^{proton}(x,Q^2;Q_0^2) = \\
& 2 \frac{\bar{\alpha} N_c}{\pi} \int^{ }_{ } \frac{d 
\gamma}{2 i \pi}
h(\gamma) \frac{v(\gamma)}{\gamma} \nonumber  
 w(\gamma) \left(\frac{Q^2}{Q_0^2}
\right)^{\gamma}
e^{\frac{\bar{\alpha} N_c}{\pi} \chi(\gamma) \ln(\frac{1}{x})}
\end{eqnarray}
where $\omega$ can
be interpreted as the Mellin transformed probability of finding an onium
of transverse mass $M^{2}$ in the proton, and $Q_{0}^{2}$ is a typically non
perturbative proton scale.
\par
We can use this generic result to get some predictions for $F_{T}$,
$F_{L}$, and $F_{G}$ (respectively the transverse, longitudinal, and
gluon structure functions) corresponding to $h_{T}$, $h_{L}$, and $h_{G}$:
\begin{eqnarray}
\label{defh}
\left(\begin{array}{c}
h_T \\ h_L \\ h_G \end{array} \right) =  & \frac{\bar{\alpha}}{ 3 \pi \gamma}
\frac{(\Gamma(1 - \gamma) \Gamma(1 + \gamma))^3}{\Gamma(2 - 2\gamma) \Gamma(2 +
2\gamma)} \frac{1}{1 - \frac{2}{3} \gamma}  . \nonumber \\
& \left( \begin{array}{c} (1 + \gamma)
(1 - \frac{\gamma}{2}) \\ \gamma(1 - \gamma) \\ 1 \end{array} \right)
\end{eqnarray}
The integral in $\gamma$ is performed by the steepest descent method. The saddle
point is at
$\gamma_{C} = \frac{1}{2} (1 -a \ln \frac{Q}{Q_{0}} )$
where
$a=\left(\frac{\bar{\alpha} N_c}{\pi} 7 \zeta(3) \ln\frac{1}{x}\right) ^{-1}$
It can be shown that the considered approximation is valid when
$\ln Q/Q_{0} / \ln (1/x) << 1$, that is small $x$, moderate $Q/Q_{0}$,
kinematical domain. We finally obtain:
\begin{eqnarray}
\label{predF2}
& F_2 \equiv F_T + F_L \\
& = C a^{1/2} e^{(\alpha_{P} -1) \ln\frac{1}{x}} \frac{Q}{Q_0}
e^{- \frac{a}{2} \ln^2 \frac{Q}{Q_0}} \nonumber
\end{eqnarray}
where
$\alpha_{P} -1 = \frac{4 \bar{\alpha} N_{C} \ln 2}{\pi}$.
$C$, $\alpha_{P}$ and $Q_{0}$ will be taken as free parameters for the fit
of the H1 data, it will then be possible to compare with the values of
$\alpha_{P}$ predicted by theory. We get finally $R$, and $F_{G}/F_{2}$,
which are independent of the overall normalisation $C$:
\begin{eqnarray}
\label{fgf2}
& \frac{F_G}{F_2}  =  \frac{1}{h_T + h_L} |_{\gamma = \gamma_c} \\
& \equiv \frac{3
\pi \gamma_c}{\bar{\alpha}} \frac{1 - \frac{2}{3}\gamma_c}{1 + \frac{3 }{2}
\gamma_c - \frac{3}{2}\gamma_c^2} \frac{\Gamma(2 - 2 \gamma_c) 
\Gamma(2 + 2 \gamma_c)}
{(\Gamma(1 - \gamma_c)\Gamma(1 + \gamma_c))^3} \nonumber
\end{eqnarray}
\begin{eqnarray}
R  = \frac{h_L}{h_T}(\gamma_c) = \frac{\gamma_c (1 - \gamma_c)}{(1 + \gamma_c)(1
 - \frac{\gamma_c}{2})}
\end{eqnarray}
where $\gamma_{C}$ is the saddle point value.

\section{$F_{2}$ fit and prediction for $F_{G}$ and $R$}
In order to test the accuracy of the $F_{2}$ parametrisation obtained in formula
(4), a fit using the recently published data from the H1 experiment
\cite{H1} has been
performed \cite{ourpap}.
We have only used the points with $Q^{2} \leq 150 GeV^{2}$ to remain
in the domain of validity of the QCD dipole model. The $\chi^{2}$ is 101
for 130 points, and the values of the parameters are $\alpha_{P}=1.282$,
$Q_{0}=0.63 GeV$, $C=0.077$. The result of the fit is shown in figure 1a.
The corresponding effective coupling constant to the obtained
value of $\alpha_{P}$ is $\bar{\alpha} =0.11$, close
to $\alpha (M_{Z})$ used in the H1 QCD fit. The value of $Q_{0}$ corresponds
to a tranverse size of 0.3 fm which is the expected non perturbative 
scale.
\par
Relation (5) provides a paramater-free
prediction for the gluon density (not shown in the figure) which is in good
agreement with the results obtained by the H1 QCD fits based on a NLO DGLAP
evolution equation \cite{ourpap}.
We also give a prediction for the value of $R$,
which is given in figure 1b. The only parameters which enters this prediction
is $Q_{0}$, determined by the $F_{2}$ fit.
The corresponding curve (full line) is compared with the one loop
approximation of the $h$ functions (dashed curve) of formula (3).
The comparison of the two curves exhibits
the ($\ln 1/x$) terms resummation effects
on the coefficient functions $h_T$ and $h_L$. The
measurement of $R$ might be an opportunity to distinguish between the BFKL
and DGLAP mechanisms, R being expected to be higher with the DGLAP
mechanism. We thus think that a measurement of $R$ in this region would
be useful.

\section{Hard diffraction in the QCD dipole model:}
The success of the dipole model applied to the proton structure function
suggests to extend the investigations to other inclusive processes, 
in particular
to diffractive dissociation. We can distinguish two different components:
\newline
- the "elastic" term which represents the elastic scattering of the onium
on the target proton
\newline
- the "triple-pomeron" term which represents the sum of all dipole-dipole
interactions (it is dominant at large masses of the excited system).
\newline
Let us describe in more details each of the two components. The 
"triple-pomeron"
term dominates at low $\beta$, where $\beta$ is the ratio between
Bjorken-$x$ and $xp$, the proton momentum fraction carried by the 
"pomeron" \cite{inelcomp}.
This component, integrated over $t$, the momentum transfer, is factorisable
in a part depending only on $xp$ (flux factor) and on a part depending only on
$\beta$ and $Q^{2}$ ("pomeron" structure function) \cite{inelcomp}.
\begin{eqnarray}
F_{2}^{D(3)} (Q^{2},xp,\beta)  = \Phi (xp) F_{P}(Q^{2},\beta) \\
\Phi (xp)   = 8 xp^{1-2 \alpha_{P}}   \left(
\bar{\alpha} N_c 7 \zeta(3) \ln\frac{1}{xp}\right) ^{-3} 
\end{eqnarray}
The effective exponent (the slope of $ln F_{2}^{D}$ in $ln xp$) is
found to be dependent on $xp$ and much smaller than the BFKL exponent 
due to the $ln (xp)$ in the flux factor. $F$ is equivalent to the 
proton structure function at small $\beta$.
\par
The elastic component behaves quite differently \cite{elcomp}. 
First it dominates at
$\beta \sim 1$. It is also factorisable like the inelastic component, but
with a different flux factor, which means that the sum of the two
components will not be factorisable. The $\beta$ dependence is quite 
flat at large $\beta$, and this is due to the fact that there is an interplay
between the longitudinal and transverse components. The sum remains almost
independent of $\beta$, whereas the $R=F_{L}/F_{T}$ ratio is strongly
$\beta$ dependent. Once more, a $R$ measurement now in 
diffractive processes will be an
interesting way to distinguish the different models, as the dipole model
predicts different $\beta$ and $Q^{2}$ behaviours from the others 
\cite{McDerm}.
\par
The sum of the two components shown in figure 2 describes quite well the
H1 data. There is no further fit of the data as we chose to take the 
different parameters ($Q_{0}, \alpha_{P}$, and the normalisation factor)
from the $F_{2}$ fit. The most striking point is that we describe quite
well the factorisation breaking due to the resummation of the two components
at low and large $\beta$ \cite{papdiff}. The full line is the sum of the
two components, the dashed line the inelastic one (which dominates at low
$\beta$), and the dotted line the elastic one (which dominates at
high $\beta$).




\begin{figure}
\psfig{figure=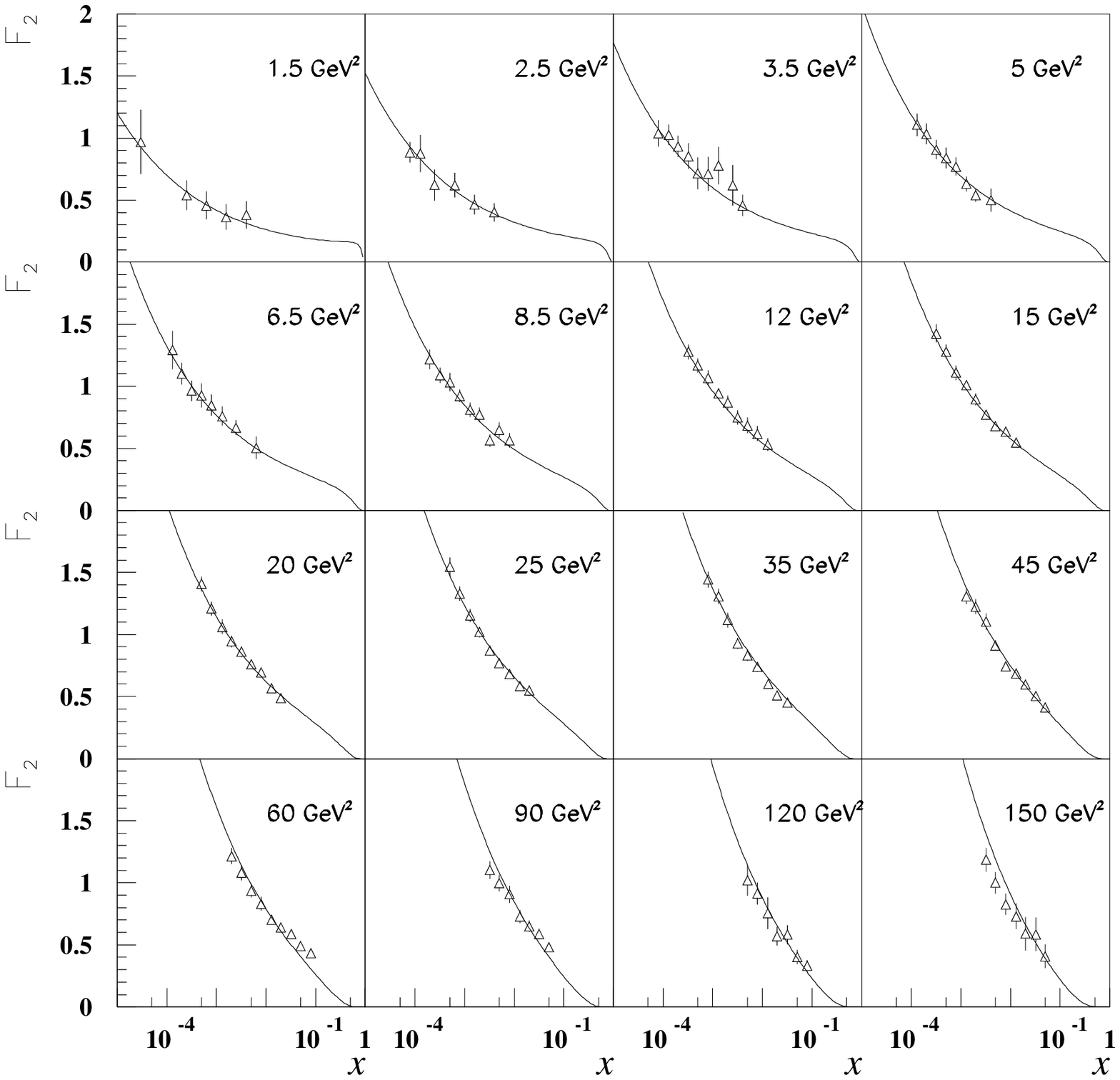,height=3in}
\psfig{figure=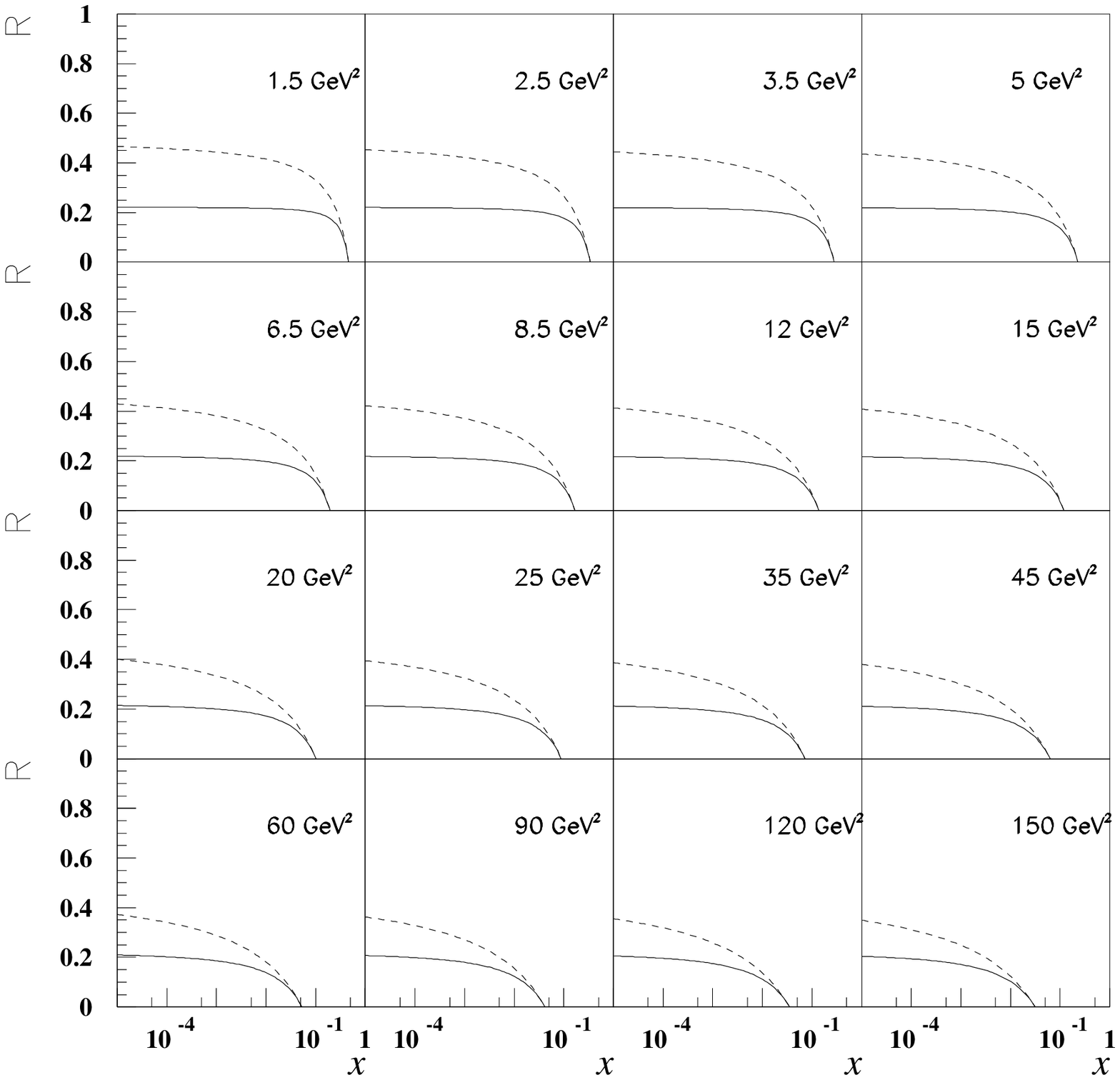,height=3in}
\caption{a: Results of the 3-parameter fit of the H1 proton
structure function for $Q^{2} \leq 150 GeV^{2}$ - b: Predictions on R
(continuous line : resummed prediction)
\label{fig:radish}}
\end{figure}



\vspace{-4mm}

\section*{Acknowledgments}
\vspace{-4mm}
The results described in the present contribution come from a fruitful
collaboration with A.Bialas, H.Navelet, R.Peschanski and S.Wallon.

\newpage

\begin{figure}
\vskip 3cm
\psfig{figure=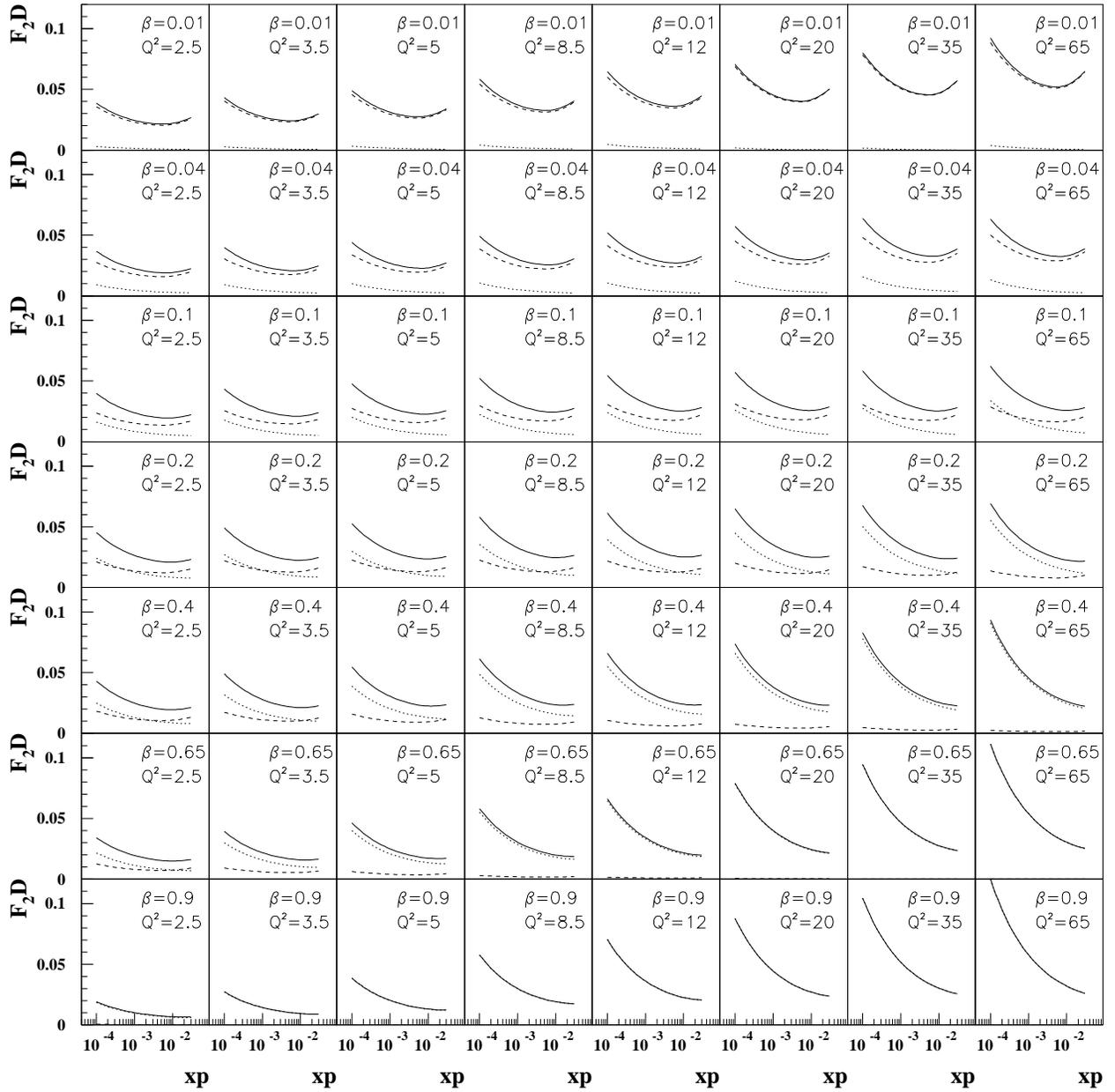,height=7in}
\caption{Prediction on F2D
(cf text) $~~~~~~~~~~~~~~~~~~~~~~~~~~~~~~~~~~~$
\label{fig:rad3}}
\end{figure}


\vspace{-4mm}

\section*{References}
\begin{thebibliography}{99}
\bibitem{Muel+}A.H.Mueller and B.Patel, {\it Nucl. Phys.} {\bf B425} (1994) 
471.,
A.H.Mueller, {\it Nucl. Phys.} {\bf B437} (1995) 107.,
N.N.Nikolaev and B.G.Zakharov, {\it Zeit. f\"ur. Phys.} {\bf C49} (1991) 607.,
A.H.Mueller, {\it Nucl. Phys.} {\bf B415} (1994) 373.

\bibitem{equ} G. Altarelli and G. Parisi, {\it Nucl. Phys.} {\bf B126} (1977)
298; V.N.
Gribov and L.N. Lipatov, {\it Sov. Journ. Nucl. Phys.} {\bf 15} (1972) 438 and
675., V.S.Fadin et al.   {\it Phys. Lett.} {\bf B60} (1975)
50; I.I.Balitsky and L.N.Lipatov, {\it Sov.J.Nucl.Phys.} {\bf 28} (1978) 822.
\bibitem{catani} S.Catani, M.Ciafaloni and Hautmann, {\it Phys. Lett.} 
{\bf B242}
 (1990) 97;
{\it Nucl. Phys.} {\bf B366} (1991) 135; J.C.Collins and R.K.Ellis, {\it Nucl.
Phys.} {\bf B360} (1991) 3; S.Catani and Hautmann, {\it Phys. Lett.} {\bf B315}
(1993) 157;
{\it Nucl. Phys.} {\bf B427} (1994) 475
\bibitem{H1} H1 coll., S.Aid et al. preprint DESY 96-039, March 1996
\bibitem{ourpap} H.Navelet, R.Peschanski, Ch.Royon, and S.Wallon,
DESY preprint 96-108, to be published in {\it Phys.Lett.B},
hep-ph/9605389, and references therein,
H.Navelet, R.Peschanski, Ch.Royon, {\it Phys.Lett.}, {\bf B366},
(1996) 329.
\bibitem{inelcomp} A.Bialas, R.Peschanski, {\it Phys. Lett.} {\bf B378}
(1996) 302 and references therein, 
A.Bialas, 2nd Cracow epiphany conference (1996)
\bibitem{elcomp} A.Bialas, R.Peschanski, preprint hep-ph/9605298,
subm. to {\it Phys. Lett.}, and references therein
\bibitem{papdiff} A.Bialas, R.Peschanski, C.Royon, to appear
in {\it Phys. Lett.}
\bibitem{McDerm} M.F.McDermott et al., Proceedings of the HERA workshop (1996)
and references therein
\end{thebibliography}
\end{document}